\documentstyle[12pt,epsfig]{article} 
 \newcommand{\be}{\begin{eqnarray}}
 \newcommand{\ee}{\end{eqnarray}}
 \newcommand{\beq}{\begin{equation}}
 \newcommand{\eeq}{\end{equation}}
 \newcommand{\ba}{\begin{array}{1}}
 \newcommand{\ea}{\end{array}}
 \newcommand{\bb}{\begin{thebibliography}}
 \newcommand{\eb}{\end{thebibliography}}

 \newcommand{\abstitle}[1]{{\small {\bf #1}}}
 \newcommand{\absauthor}[1]{{\small {\bf #1}}}
 \newcommand{\address}[1]{{\it #1}}
 
\begin{document}
 \begin{center}
%\begin{frontmatter}
%\abstitle
\abstitle{\large\bf Nucleon structure and hard p-p processes at high energies} \\
%\\ in proton with ATLAS \par }
 \vspace{0.4cm}
\absauthor{G.I.Lykasov, I.V.Bednyakov, M.A. Demichev, Yu.Yu. Stepanenko}\\
 \vspace{0.4cm}
\address{JINR, Dubna, 141980, Moscow region, Russia}
 \vspace{0.4cm}
%\begin{center}

{\bf Abstract}
\end{center}
%% Text of abstract
The production of heavy flavour hadrons in $pp$ collisions at large values of 
their transverse momenta can be a new unique source for estimation 
of intrinsic heavy quark contribution to the proton. We analyze the inclusive
production of the open strangeness and the semi-inclusive hard processes of the photon
and vector boson production accompanied by the $c$- or $b$-jets in $pp$ collisions.
We show that one should select the parton-level (sub)processes 
(and final-state signatures) that are the most sensitive to the intrinsic heavy quark 
contributions.  
We present some predictions for these processes made within 
the perturbative QCD including the intrinsic strangeness and intrinsic charm in the 
proton that can be verified in the NA61 experiment and at LHC.
%\end{abstract}
%\end{center}
%\begin{keyword}
%% keywords here, in the form: keyword \sep keyword
%charm \sep bottom \sep strangeness \sep intrinsic quarks \sep heavy flavour jets \sep hard p-p collisions
%% MSC codes here, in the form: \MSC code \sep code
%% or \MSC[2008] code \sep code (2000 is the default)

%\end{keyword}

%\end{frontmatter}

%%
%% Start line numbering here if you want
%%
% \linenumbers

%% main text
\section{Intrinsic heavy flavours in the proton}
%\label{I}
      The NA61 (CERN), CBM (Darmstadt) and NICA (Dubna) experiments can be a useful laboratory for 
      investigation of the unique structure of the proton, in particular for 
      the study of the parton distribution functions (PDFs) with high accuracy.
      It is well known that the precise knowledge of these PDFs is very important 
      for verification of the Standard Model and search for New Physics. 

      By definition, the PDF $f_a(x,\mu)$ is a function of the proton momentum fraction $x$ 
      carried by parton $a$ (quark $q$ or gluon $g$) at the QCD momentum transfer scale $\mu$. 
      For small values of $\mu$, corresponding to the long distance scales less than $1/\mu_0$, 
      the PDF cannot be calculated from the first principles of QCD 
      (although some progress in this direction 
      has been recently achieved within the lattice methods 
\cite{LATTICE}). 
      The PDF $f_a(x,\mu)$ at $\mu>\mu_0$ can be calculated by 
      solving the perturbative QCD evolution equations (DGLAP) 
\cite{DGLAP}.  
      The unknown (input for the evolution) functions $f_a(x,\mu_0)$ 
      can usually be found empirically from some 
      ``QCD global analysis'' 
\cite{QCD_anal1,QCD_anal2} of a large variety of data, typically at $\mu>\mu_0$. 

     In general, almost all $pp$ processes that took place at the LHC energies, 
     including the Higgs boson production,
     are sensitive to the charm $f_c(x,\mu)$ or bottom $f_b(x,\mu)$ PDFs. 
     Nevertheless, within the global analysis 
     the charm content of the proton at $\mu\sim\mu_c$ and 
     the bottom one at $\mu\sim\mu_b$ are both assumed to be negligible.
     Here $\mu_c$ and $\mu_b$ are typical energy scales relevant to the 
     $c$- and $b$-quark 
     QCD excitation in the proton.
     These heavy quark components arise in the proton only perturbatively
     with increasing $Q^2$-scale 
     through the gluon splitting in the DGLAP $Q^2$ evolution 
\cite{DGLAP}. 
     Direct measurement of the open charm and open bottom production 
     in the deep inelastic processes (DIS) confirms the perturbative 
     origin of heavy quark flavours 
\cite{H1:2005}. 
     However, the description of these experimental data is 
     not sensitive to the heavy quark distributions at relatively 
     large $x$ ($x>0.1$). 

\begin{figure}[h!]
\begin{center}
%\centerline{\includegraphics[width=0.50\textwidth]{nucleon.ps}}
\epsfig{file=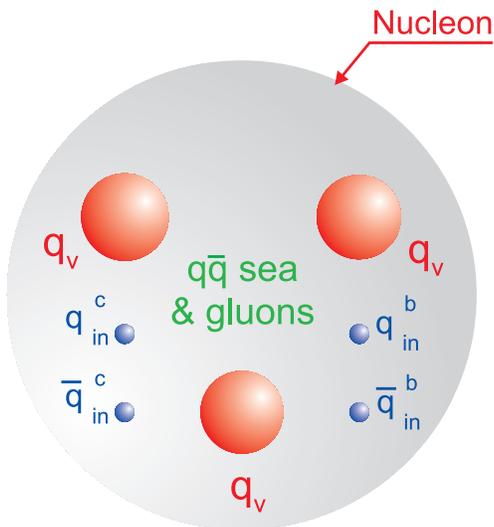,width=0.5\linewidth}
\caption{Schematic presentation of a nucleon consisting of three valence quarks
         q$_{\rm v}$, quark-antiquark q${\bar {\rm q}}$ and gluon sea, 
	 and pairs of the intrinsic charm 
	 (q$_{\rm in}^{\rm c}{\bar {\rm q}_{\rm in}^{\rm c}}$) 
	 and intrinsic bottom quarks 
	 (q$_{\rm in}^{\rm b}{\bar {\rm q}_{\rm in}^{\rm b}}$). 
}
\label{Fig_IQ}
\end{center}
\end{figure}

     As was assumed by Brodsky with coauthors in 
\cite{Brodsky:1980pb, Brodsky:1981}, 
     there are {\it extrinsic} and {\it intrinsic} 
     contributions to the quark-gluon structure of the proton. 
     {\it Extrinsic} (or ordinary) quarks and gluons are generated on 
     a short time scale associated with a large-transverse-momentum processes.
     Their distribution functions satisfy the standard QCD evolution equations. 
     {\it Intrinsic} quarks and gluons exist
     over a time scale which is independent of any probe momentum transfer. 
     They can be associated with bound-state 
{(zero-momentum transfer regime)} hadron dynamics and 
     are believed to be of nonperturarbative origin.
Figure~\ref{Fig_IQ}  gives %illustrates the 
     a schematic view %presentation 
     of a nucleon, which consists of three valence quarks q$_{\rm v}$, 
     quark-antiquark q${\bar {\rm q}}$ and gluon sea, and, for example,  
     pairs of the {\it intrinsic} charm 
(q$_{\rm in}^{\rm c}{\bar {\rm q}_{\rm in}^{\rm c}}$) and 
     {\it intrinsic} bottom quarks 
(q$_{\rm in}^{\rm b}{\bar {\rm q}_{\rm in}^{\rm b}}$). 
%%%%%%%%%%%%%%%%%%%%%%%%%%%%%%%%%%%%%%%%%%%%%%%%%%%%%%%%%%%%%%%f%%%%%%%%%%%%%%%%%%

%%%%%%%%%%%%%%%%%%%%%%%%%%%%%%%%%%%%%%%%%%%%%%%%%%%%%%%%%%%%%%%%%%%%%

       It was shown in 
\cite{Brodsky:1981}
       that the existence of {\it intrinsic} heavy quark pairs 
       $c{\bar c}$ and $b{\bar b}$ within the proton state 
       could be due to the virtue of gluon-exchange and vacuum-polarization graphs. 
       On this basis, %Using such a mechanism 
       within the MIT bag model 
\cite{Golowich:1981}, 
       the probability to find the five-quark component 
       $|uudc{\bar c}\rangle$ bound within the nucleon bag 
       was estimated to be about 1--2\%. 

%       The idea of the intrinsic charm existence in the 
%       proton was first put forward thirty years ago by 
       Initially in 
\cite{ Brodsky:1980pb,Brodsky:1981}
       S.Brodsky with coauthors %      They 
       have proposed %assumed 
       existence of the 5-quark state $|uudc{\bar c}\rangle$ 
       in the proton 
(Fig.~\ref{Fig_IQ}). 
       Later some other models were developed. 
       One of them considered a quasi-two-body state 
       ${\bar D}^0(u{\bar c})\, {\bar\Lambda}_c^+(udc)$ in the proton %see for example
\cite{Pumplin:2005yf}. % and references therein.   
       In 
\cite{Pumplin:2005yf}--\cite{Nadolsky:2008zw} 
%\cite{Pumplin:2005yf,Pumplin:2007wg,Nadolsky:2008zw}  
       the probability to find the intrinsic charm (IC) in the proton 
       (the weight of the relevant Fock state in the proton)
       was assumed to be 1--3.5\%. 
       The probability of the intrinsic bottom (IB) in the proton 
       is suppressed by the factor $m^2_c/m^2_b\simeq 0.1$ 
\cite{Polyakov:1998rb}, where $m_c$ and $m_b$ are the masses of 
       the charmed and bottom quarks. 
       Nevertheless, it was %also 
       shown that the IC %intrinsic charm 
       could result in a sizable contribution 
       to the forward charmed meson production
\cite{Goncalves:2008sw}. 
       Furthermore the IC ``signal'' can 
       constitute almost %reach about 
       100\% %in 
       of the inclusive spectrum of $D$-mesons produced at 
       high pseudorapidities $\eta$ 
       and large transverse momenta $p_T$
       in $pp$ collisions at LHC energies 
\cite{LBPZ:2012}. 

       If the distributions of the intrinsic charm or bottom in the 
       proton are hard enough and are similar in the shape to the valence quark distributions
       (have the valence-like form),  
       then the production of the charmed (bottom) mesons or charmed (bottom) 
       baryons in the fragmentation region should be similar 
       to the production of pions or nucleons. 
       However, the yield of this production depends on the probability to find the 
       intrinsic charm or bottom in the proton, but this yield looks too small.     
       The PDF which included the IC contribution in the proton 
       have already been used in the perturbative QCD calculations in  
\cite{Pumplin:2005yf}-\cite{Nadolsky:2008zw}.

\begin{figure}[h!]
\begin{center}
\epsfig{file=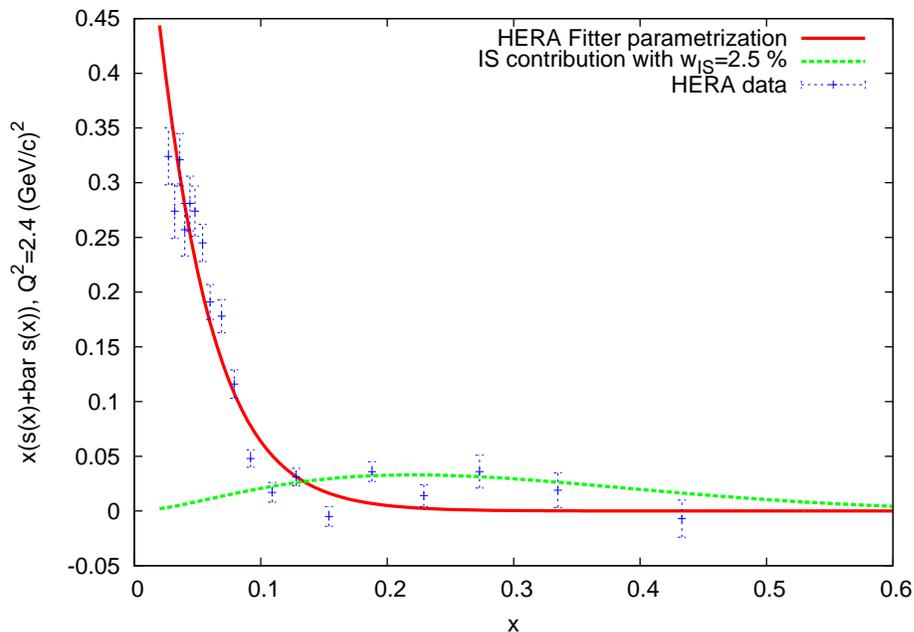,width=0.8\linewidth}
%\centerline{\includegraphics[width=0.50\textwidth]{Sx_IS.ps}}
 \caption{The distributions of strange quarks $xS(x)=x(s(x)+{\bar s}(x))$ in the proton; the
solid line is the HERA Fitter parametrization of of $xS(x)$ at $Q^2=$2.4 GeV$/$c, the dashed curve 
is the contribution of the {\it intrinsic} strangeness (IS) in the proton with the probability 
2.5 \%. The HERA data were taken from \cite{HERMES:2008}.
}
\label{Fig_2IS}
\end{center}
\end{figure}
%%%%%%%%%%%%%%%%%%%%%%%%%%%%%%%%%%%%%%%%%%%%%%%%%%%%%%%%%%%%%%%%%%%%%%%%%%%%%%%%%%%%%%%%%%%%%%%%%%%%%%
%Assuming 
   Due to the nonperturbative {\it intrinsic} heavy quark components one can expect 
   some excess of the heavy quark PDFs over 
   the ordinary sea quark PDFs at $x>0.1$. 
   The ``signal'' of these components can be visible in the observables of 
   the heavy flavour production in semi-inclusive $ep$ DIS and inclusive 
   $pp$ collisions at high energies. 
   For example, it was recently shown that rather good description of the HERMES data 
   on the $xf_s(x,Q^2)+xf_{\bar s}(x,Q^2)$ at $x>0.1$ and 
   $Q^2=2.5$~GeV$/c^2$
\cite{IS:2012,HERMES:2008} could be achieved due to existence of {\it intrinsic} strangeness in the proton, 
   see Fig.~\ref{Fig_2IS}. One can see from Fig.~\ref{Fig_2IS} that the inclusion of the intrinsic
   strangeness allows us to describe the HERA data rather satisfactorily in the whole 
   $x$-region both at
   $x\leq$ 0.1 and $x>$ 0.1. The new HERMES data confirm that the $x$-dependence of 
   $xS(x)$ at $Q^2=$ 1.9 (GeV$/$c)$^2$ is not decreasing at $x>$0.1, see the presentation 
   at the workshop DIS2013 \cite{DIS:2013}.     

   Similarly, possible existence of the intrinsic charm in the proton
   can lead to some enhancement in the inclusive spectra of the open charm hadrons, 
   in particular $D$-mesons, produced at the LHC in $pp$-collisions 
   at high pseudorapidities $\eta$
   and large transverse momenta $p_T$ 
\cite{LBPZ:2012}.

%%%%%%%%%%%%%%%%%%%%%%%%%%%%%%%%%%%%%%%%%%%%%%%%%%%%%%%%%%%%%%%%%%%%%%%%%%%%%%%%%%%%%%%%%%%%%%%%%%%%%%%%%%
       The probability distribution for the 5-quark state ($|uudc{\bar c}\rangle$) 
       in the light-cone description of the proton was first calculated in 
\cite{Brodsky:1980pb}. 
       The general form for this distribution calculated within the light-cone dynamics
       in the so-called BHPS model 
\cite{Brodsky:1980pb,Brodsky:1981} 
      can be written as 
\cite{IS:2012} 
\begin{eqnarray}
P(x_1,{\dots},x_5)=N_5\delta\left(1-\sum_{j=1}^5x_j\right)\times \\
\nonumber
\times\left(m_p^2-\sum_{j=1}^5\frac{m_j^2}{x_j}\right)^{-2},
\label{def:B}
\end{eqnarray}
       where $x_j$ is the momentum fraction of the parton, $m_j$ is its mass and $m_p$ is the
       proton mass.  
       Neglecting the light quark ($u,d, s$) masses and the proton
       mass in comparison to the $c$-quark mass and integrating 
(\ref{def:B}) 
       over $dx_1...dx_4$ one can get the probability to find the  
       intrinsic charm with momentum fraction $x_5$ in the proton \cite{Peng_Chang:2012}: 
\begin{eqnarray}
P(x_5) &=& \frac{1}{2}{\tilde N}_5x_5^2 \Big[ \frac{1}{3}(1-x_5)(1+10x_5+x_5^2)- \nonumber \\ 
      && 2x_5(1+x_5)\ln(x_5) \Big],
\label{def:fcPumpl}
\end{eqnarray}
     where ${\tilde N}_5=N_5/m^4_{4,5}, m_{4,5}=m_c=m_{\bar c}$, 
     the normalization constant $N_5$ determines some 
     probability $w^{}_{\rm IC}$ to find %, in particular, 
     the Fock state $|uudc{\bar c}\rangle$ in the proton.
%%%%%%%%%%%%%%%%%%%%%%%%%%%%%%%%%%%%%%%%%%%%%%%%%%%%%%%%%%%%%%%%%%%%%%%%%%%%%%%%%%%%%%%%%%%%%%%%%%%%%%%%%%
%%%%%%%%%%%%%%%%%%%%%%%%%%%%%%%%%%%%%%%%%%%%%%%%%%%%%%%%%%%%%%%%%%%%%%%%%%%%%%%%%%%%%%%%%%%%%%%%%%%%%%%%%%

\begin{figure}[h!]
\begin{center}
\epsfig{file=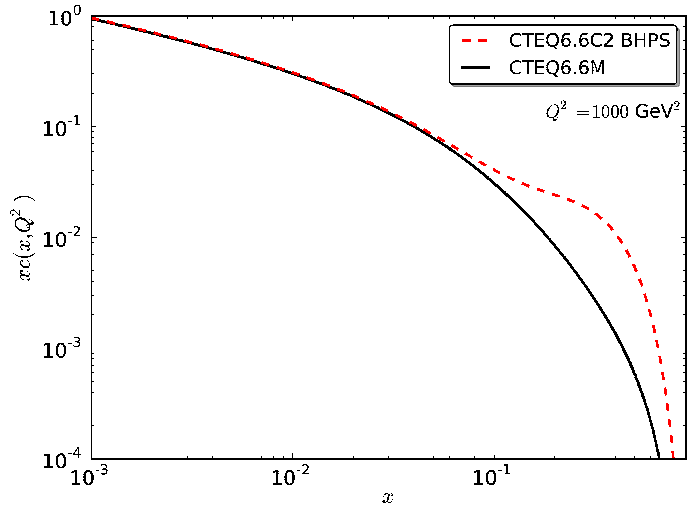,width=0.6\linewidth}
%\centerline{\includegraphics[width=0.50\textwidth]{cteq66BHPSQ21000.ps}}
 \caption{Distributions of the charm quark in the proton
  at $Q^2=$1000 GeV$^2$. 
  The solid line is the standard perturbative sea charm density distribution $xc_{\rm rg}(x)$ only, whereas 
  the dashed curve is the charm quark distribution function, 
  for the sum of the intrinsic charm density $xc_{\rm in}(x)$ 
(see (\ref{def:fcPumpl})) and $xc_{\rm rg}(x)$.
}. 
\label{Fig_2IC}
\end{center}
\end{figure}

 Figure~\ref{Fig_2IC} illustrates the IC contribution in comparison to the 
    conventional sea charm quark distribution in the proton. 

%%%%%%%%%%%%%%%%%%%%%%%%%%%%%%%%%%%%%%%%%%%
     The solid line in 
Fig.~\ref{Fig_2IC} shows the standard perturbative sea charm density distribution 
     $xc_{\rm rg}(x)$ (ordinary sea charm) in the proton from CTEQ6.6M
\cite{Nadolsky:2008zw} as a function of $x$ at    
     $Q^2=$ 1000 GeV$^2$.
     The dashed curve in
Fig.~\ref{Fig_2IC} is the sum of the intrinsic charm density $xc_{\rm in}(x)$ from 
     CTEQ6.6C2 BHPS with the IC probability $w^{}_{\rm IC}= 3.5$\% and $xc_{\rm rg}(x)$ at 
     the same $Q^2$ \cite{Nadolsky:2008zw}.    
%%%%%%%%%%%%%%%%%%%%%%%%%%%%%%%%%%%%%%%%%%%
      One can see from 
Fig.~\ref{Fig_2IC} 
      that the IC distribution (with $w^{}_{\rm IC} = 3.5$\%) given by 
(\ref{def:fcPumpl}) 
      has a rather visible enhancement at $x\sim $~0.2--0.5 and 
      this distribution is much larger ( by an order and more of magnitude) 
      than the sea (ordinary) charm density distribution in the proton. 

     As a rule, the gluons and sea quarks play the key %main 
     role in hard processes of open charm hadroproduction. 
     Simultaneously, due to the nonperturbative {\it intrinsic} heavy quark components 
     one can expect some excess of these heavy quark PDFs over 
     the ordinary sea quark PDFs at $x>0.1$. 
     Therefore the existence of this intrinsic charm component %in the proton
     can lead to some enhancement in the inclusive spectra of open charm hadrons, 
     in particular $D$-mesons, produced at the LHC in $pp$-collisions 
     at large pseudorapidities $\eta$ and large transverse momenta $p_T$ 
\cite{LBPZ:2012}.
    Furthermore, as we know from 
\cite{Brodsky:1980pb}-\cite{Nadolsky:2008zw}
%    It is also well-known that
    photons produced in association with heavy quarks $Q(\equiv c,b)$ in the final 
    state of $pp$-collisions provide valuable information about the parton 
    distributions in the proton
\cite{Pumplin:2005yf}-\cite{Thomas:1997}.

\smallskip
    In this paper, having in mind these considerations %and hints, 
    we will first discuss where the above-mentioned 
    heavy flavour Fock states in the proton could be searched for %looked 
    at high energies. Following this we analyze in detail, and give predictions for,
    the open strangeness production in $pp$ collisions and the LHC semi-inclusive 
    $pp$-production of prompt photons and vector bosons accompanied by  $c$-jets or 
    $b$-jets including the {\it intrinsic} strange or {\it intrinsic} charm component 
    in the PDF.

%%%%%%%%%%%%%%%%%%%%%%%%%%%%%%%%%%%%%%%%%%%%%%%%%%%%%%%%%%%%%%%%%%%%%%%%%%%%%%%%%%%%%%%%%%%%%%%%%%%%%%%%%%

\section{Intrinsic heavy quarks in hard $pp$ collisions}
%\label{II}
%\subsection{Intrinsic charm and beauty contribution}
\subsection{Where can one look for the intrinsic heavy quarks?}
%$\bullet~${\bf Intrinsic charm}\\
    It is known that in the $pp$ production of heavy flavour hadrons 
    at large momentum transfer the hard QCD interactions of two sea 
    quarks, two gluons and a gluon with a sea quark play the main role.
%%%%%%%%%%%%%%%%%%%%%%%%%%%%%%%%%%%%%%%%%%%%%%%%%%%%%%%%%%%%%%%%%%%%%
According to the model of hard scattering 
\cite{AVEF:1974}--\cite{FF:AKK08}
%\cite{AVEF:1974,Nasson,Nasson1,Nasson2,FF,FFF1,FFF2,Mangano:2010,PDF:CTEQ,PDF:MRST,FF:AKK08}, 
the relativistic invariant
inclusive spectrum of the hard process $p+p\rightarrow h+X$ can be related to
the elastic parton-parton subprocess $i+j\rightarrow i^\prime +j^\prime$,
where $i,j$ are the partons (quarks and gluons). 
This spectrum can be presented in the following general form 
\cite{FF}--\cite{FFF2} (see also \cite{BGLP:2011,BGLP:2012}):
\begin{eqnarray}
%\label{def:rho_c} %\rho(x,p_t)\equiv 
E\frac{d\sigma}{d^3p}= %      &=&
\sum_{i,j}\!\int\! d^2k_{iT}\!\int\! d^2k_{jT}\!\int_{x_i^{\min}}^1dx_i\!\int_{x_j^{\min}}^1dx_j\times 
\label{def:rho_c}\\
\nonumber
%\times\\ &\times&
\times f_i(x_i,k_{iT})f_j(x_j,k_{jT}) %\frac{1}{\pi}
\frac{d\sigma_{ij}({\hat s},{\hat t})}{d{\hat t}}\frac{D_{i,j}^h(z_h)}{\pi z_h}.
\label{def:hscm} 
%\nonumber
%\label{def:rho_c}
\end{eqnarray}
   Here $k_{i,j}$ and $k_{i,j}^\prime$ are the four-momenta of the partons $i$ or $j$ 
   before and after the elastic parton-parton scattering, respectively; 
   $k_{iT}, k_{jT}$ are the transverse momenta of the partons $i$ and $j$;  
   $z$ is the fraction of the hadron momentum from the parton momentum; 
   $f_{i,j}$ is the PDF; and $D_{i,j}$ is the fragmentation function (FF) 
   of the parton $i$ or $j$ into a hadron $h$.

    When %with 
    the transverse momenta of the partons are neglected 
    in comparison with the longitudinal momenta, 
    the variables ${\hat s}$, ${\hat t}$, ${\hat u}$ and $z_h$ can be 
    presented in the following forms \cite{FF}:
\begin{eqnarray}
{\hat s}=x_i x_j s,\quad {\hat t}=x_i \frac{t}{z_h}, \quad
{\hat u}=x_j \frac{u}{z_h}, 
\label{def:stuzh}\\
\nonumber
%\quad 
z_h=\frac{x_1}{x_i}+\frac{x_2}{x_j},
%\label{def:stuzh}
\end{eqnarray}
     where
\begin{eqnarray}
x_1=-\frac{u}{s}=\frac{x_T}{2}\cot({\theta}/{2}), \\
\nonumber
%\quad
x_2=-\frac{t}{s}=\frac{x_T}{2}\tan({\theta}/{2}), \\
\nonumber
%\quad
x_T=2\sqrt{t u}/s=2p_T/\sqrt{s}.
\end{eqnarray}
      Here as usual, 
      $s=(p_1+p_2)^2$,
      $t=(p_1-p_1^\prime)^2$,
      $u=(p_2-p_1^\prime)^2$, 
     and $p_1$, $p_2$, $p_1^\prime$ are the 4-momenta of the colliding protons 
     and the produced hadron $h$, respectively; 
     $\theta$ is the scattering angle for the hadron $h$ in the $pp$ c.m.s.
     The lower limits of the integration in
(\ref{def:rho_c}) are 
\begin{eqnarray}
x_i^{\min}=\frac{x_T \cot(\frac{\theta}{2})}{2-x_T \tan(\frac{\theta}{2})}, 
\label{def:xijmn}\\
\nonumber
%\qquad
x_j^{\min}=\frac{x_i x_T \tan(\frac{\theta}{2})}{2x_i-x_T \cot(\frac{\theta}{2})}.
%\label{def:xijmn}
\end{eqnarray} 
%%%%%%%%%%%%%%%%%%%%%%%%%%%%%%%%%%%%%%%%%%%%%%%%%%%%%%%%%%%%%%%%%%%%%%%%%%%%%%%%%%%%%%%%

     The lower limits of the integration in
(\ref{def:rho_c})  
     can be presented also in the following form:

\begin{eqnarray}
x_i^{\min}  = \frac{x_R+x_F}{2-(x_R-x_F)}, 
\label{def:xijmin}\\
\nonumber
%\qquad
x_j^{\min}  = \frac{x_i(x_R-x_F)}{2x_i-(x_R+x_F)}~,    
%\label{def:xijmin}
\end{eqnarray}
where 
     the Feynman variable $x_F$ of the produced hadron 
%     for example the $D$-meson,
     can be expressed via  
     the variables $p_T$ and $\eta$, or $\theta$ 
     being the hadron scattering angle in the $pp$ c.m.s: 
\begin{eqnarray}
x_F \equiv \frac{2p_{z}}{\sqrt{s}}
=\frac{2p_T}{\sqrt{s}}\frac{1}{\tan\theta}
=\frac{2p_T}{\sqrt{s}}\sinh(\eta). 
%\label{def:xFetapt}
\label{def:xFptteta}
\end{eqnarray} 
      One can see from 
(\ref{def:xijmin}) that, at least, one of the low limits $x_i^{\min}$ of the integral 
(\ref{def:rho_c}) must be $\geq x_F$. 
      Thus if $x_F\geq 0.1$, then $x_i^{\min}>0.1$, 
      where the ordinary ({\it extrinsic}) charm distribution is completely negligible
      in comparison with the {\it intrinsic} charm distribution. 
      Therefore, at $x_F\geq 0.1$, or equivalently %consequently
      at the charm momentum fraction $x_c> 0.1$ 
      the {\it intrinsic} charm distribution intensifies the charm %sea 
      PDF contribution into charm hadroproduction substantially
(see Fig.~\ref{Fig_2IC}). % in a few order of magnitude.
      As a result, %It can increase 
      the spectrum of the open charm hadroproduction can be increased 
      in a certain region of $p_T$ and $\eta$ (which corresponds 
      to $x_F\geq 0.1$ in accordance to (\ref{def:xijmin})). 
      We stress that this excess (or even the very possibility to observe relevant events in this 
      region) is due to the non-zero contribution of IC component at $x_c > x_F> 0.1$ 
      (where non-IC component completely vanishes).

      This possibility was demonstrated for 
      the $D$-meson production at the LHC in 
\cite{LBPZ:2012}.     
       It was shown that the $p_T$
       spectrum of $D$-mesons is enhanced at pseudorapidities of $3<\eta<5.5$ and 
       10 GeV$/c<p_T<$ 25 GeV/$c$ due to the IC contribution, which was included using the
       CTEQ66c PDF \cite{Nadolsky:2008zw}. For example, due to the IC PDF, with probability about 
       3.5 $\%$, the $p_T$-spectrum increases by a factor of 2 at $\eta=4.5$.
       A similar  effect was predicted in \cite{Kniehl:2012ti}.   

      One expects a similar enhancement in 
      the experimental spectra %distributions 
      of the open bottom production 
      due to the (hidden) intrinsic bottom (IB) in the proton, which could have  
      a distribution very similar to the one given in
  (\ref{def:fcPumpl}). 
      However, the probability $w_{\rm IB}$ to find
      the Fock state with the IB contribution $|uudb{\bar b}\rangle$  in the proton 
      is about 10 times smaller than the IC probability $w_{\rm IC}$ 
      due to relation $w_{\rm IB}/w_{\rm IC}\sim m_c^2/m_b^2$ 
%      where $m_b$ is the bottom quark mass 
\cite{Brodsky:1981,Polyakov:1998rb}.
     
     The IC ``signal'' can be studied not only in the inclusive open (forward) 
     charm hadroproduction at the LHC, but also in some other processes, such as
     production of real prompt photons $\gamma$ or virtual ones $\gamma^*$, or $Z^0$-bosons
     (decaying into dileptons) accompanied by $c$-jets in the kinematics available to the ATLAS and CMS
     experiments.
     The contributions of the heavy quark states in the proton could be 
     investigated also in the $c(b)$-jet 
     production accompanied by the vector bosons $W^\pm,Z^0$. 
     Similar kinematics given by (\ref{def:stuzh}-\ref{def:xFptteta}) can also be applied to 
     these hard processes.

%%%%%%%%%%%%%%%%%%%%%%%%%%%%%%%%%%%%%%%%%%%%%%%%%%%%%%%%%%%%%%%%%%%%%%%%%%%%%%%%%%%%%%%%
Actually, the parton distribution functions $f_i(x_i,k_{iT})$ also depend on  
the four-momentum transfer squared $Q^2$ that is related to the Mandelstam variables
${\hat s},{\hat t},{\hat u}$ for the elastic parton-parton scattering \cite{FFF2} 
\begin{eqnarray}
Q^2~=~\frac{2{\hat s}{\hat t}{\hat u}}{{\hat s}^2+{\hat t}^2+{\hat u}^2}
\label{def:Qsqr}
\end{eqnarray} 
%%%%%%%%%%%%%%%%%%%%%%%%%%%

    Calculating spectra by Eq.(\ref{def:rho_c}) we used 
    the PDF which includes the IS (and does not include it) \cite{Nadolsky:2008zw},
%CTEQ66c which includes the IC contribution and CTEQ66 that does not \cite{Nadolsky:2008zw}
%    include the IC
     the FF of the type AKK08 
\cite{FF:AKK08}  and 
    $d\sigma_{ij}({\hat s},{\hat t})/d{\hat t}$ 
    calculated within the LO QCD and presented, for example, in %obtained within LO QCD, see, for example 
\cite{Mangano:2010}.  

%%%%%%%%%%%%%%%%%%%%%%%%%%%%%%%%%%%%%%%%%%%%%%%%%%%%%%%%%%%%%%%%%%%%%
     One can see from Eq.(\ref{def:xFptteta}) that the Feynman variable $x_F$ of the produced hadron
     can be expressed via  %is related to 
     the variables $p_T$ and $\eta$, or $\theta$ %%%%% $\theta^*$ 
     the hadron scattering angle in the $pp$ c.m.s, 
     At small scattering angles of the produced hadron Eq.(\ref{def:xFptteta}) becomes 
\begin{eqnarray}
x_F\sim \frac{2p_T}{\sqrt{s}}\frac{1}{\theta}.
\label{def:xFtetapt}
\end{eqnarray} 
       It is clear that for fixed $p_T$ an outgoing hadron must possess a very small $\theta$ or very large $\eta$
        in order to have large $x_F$ (to follow forward, or backward direction).

       In the fragmentation region (of large $x_F$) the Feynman variable $x_F$ 
       of the produced hadron is related to 
       the variable $x$ of the intrinsic charm quark in the proton, and  
       according to the longitudinal momentum conservation law, 
       the $x_F \simeq x$ (and $x_F < x$). %it can be not so far from $x$. 
       Therefore, the visible excess of the inclusive spectrum, for example, of $K$-mesons
       can be due to the enhancement of the IS distribution 
(see Fig.~\ref{Fig_2IS}) at $x>$ 0.1.

\section{Intrinsic strangeness}
%\label{III}
\subsection{Open strangeness production in hard $pp$ collisions}
%$\bullet~${\bf Intrinsic strangeness}\\
Let us analyze now how the possible existence of the intrinsic strangeness in the proton 
can be visible in $pp$ collisions. For example, consider the $K^-$-meson production in 
the process $pp\rightarrow K^-+X$. Considering the intrinsic strangeness in the proton \cite{IS:2012}
we calculated the inclusive spectrum $ED\sigma/d^3p$
of such mesons within the hard scattering model (Eq.(\ref{def:hscm})), 
which describes satisfactorily the
HERA and HERMES data on the DIS. The FF and the parton cross sections were taken from
\cite{FF:AKK08,Mangano:2010}, respectively, as mentioned above.
\begin{figure}[h!]
\begin{center}
\epsfig{file=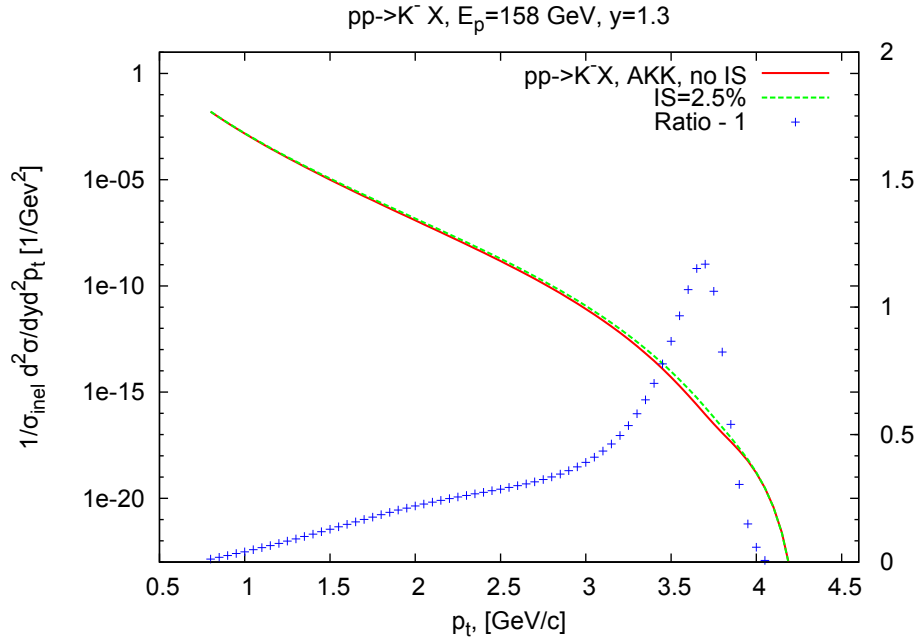,width=0.8\linewidth}
%\centerline{\includegraphics[width=0.50\textwidth]{IS_NA61_1_3_v1.ps}}
 \caption{The $K^-$-meson distributions (with and without intrinsic strangeness contribution) 
  over the transverse momentum $p_t$ for $pp\rightarrow K^- + X$  
  at the initial energy $E=$ 158 GeV, the rapidity $y=$1.3 and $p_t\geq$ 0.8 GeV$/$c.
} 
\label{Fig_5}
\end{center}
\end{figure}
\begin{figure}[h!]
\begin{center}
\epsfig{file=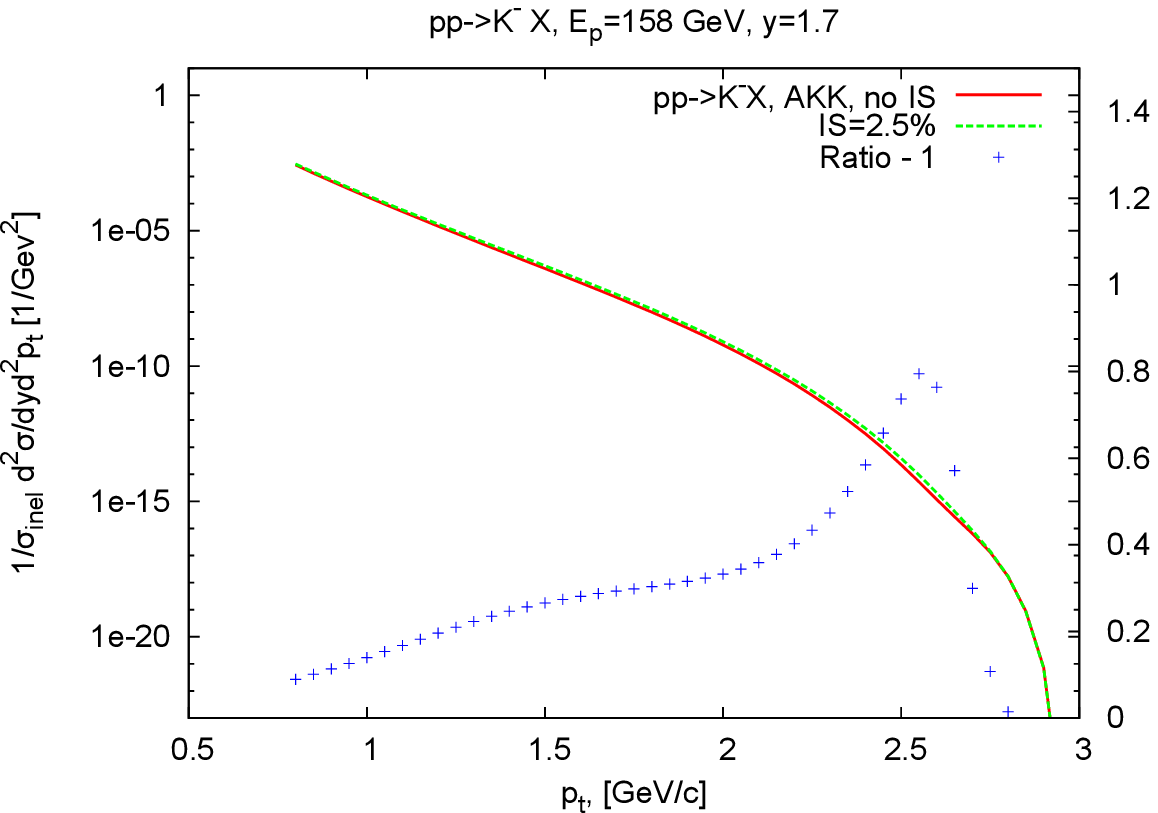,width=0.8\linewidth}
%\centerline{\includegraphics[width=0.50\textwidth]{IS_NA61_1_7_v1.pdf}}
 \caption{The $K^-$-meson distributions (with and without intrinsic strangeness contribution) 
  over the transverse momentum $p_t$ for $pp\rightarrow K^- + X$  
  at the initial energy $E=$ 158 GeV, the rapidity $y=$1.7 and $p_t\geq$ 0.8 GeV$/$c.
} 
\label{Fig_6}
\end{center}
\end{figure}
In Figs.~(\ref{Fig_5},\ref{Fig_6}) the inclusive $p_t$-spectra of $K^-$-mesons produced in $pp$
collision at the initial energy $E_p=$158 GeV are presented at the rapidity $y=$1.3 (Fig~\ref{Fig_5})
and $y=$1.7 (Fig~\ref{Fig_6}). The solid lines in  Figs.~(\ref{Fig_5},\ref{Fig_6}) correspond to our
calculation ignoring the {\it intrinsic strangeness} (IS) in the proton and the dashed curves correspond 
to the calculation including the IS with the probability about 2.5$\%$, according to \cite{IS:2012}. 
The crosses show the ratio of our calculation with the IS and without the IS minus 1.
One can see from Figs.~(\ref{Fig_5},\ref{Fig_6}), right axis, that the IS signal can be above 200 $\%$ 
at $y=$ 1.3, $p_t=$ 3.6-3.7 Gev$/$c and slightly smaller, than 200 $\%$ at 
$y=$ 1.7, $p_t\simeq$ 2.5 Gev$/$c. Actually, this is our prediction for the NA61 experiment that
is now under way at CERN.  
%%%%%%%%%%%%%%%%%%%%%%%%%%%%%%%%%%%%%%%%%%%%%%%%%%%%%%%%%%%%%%%%%%%%%%%%%%%%%%%%%%%%%%%%%%%%%%%
%%%%%%%%%%%%%%%%%%%%%%%%%%%%                             %%%%%%%%%%%%%%%%%%%%%%%%%%%%%%%%%%%%%%

\section{Intrinsic charm}
%\label{IV}
\subsection{Prompt photon and $c$-jet production} 

    Recently 
    the investigation of prompt photon and $c(b)$-jet production in
    $p{\bar p}$ collisions at $\sqrt{s}=1.96$~TeV was carried out at the TEVATRON 
\cite{D0:2009}-\cite{Aaltonen:2009wc}.
    In particular,
    it was observed that the ratio of the experimental spectrum of the prompt photons, 
    (accompanied by the $c$-jets) to the relevant theoretical expectation 
    (based on the conventional PDF which ignored the {\it intrinsic} charm) 
    increases with $p_T^\gamma$ up to factor about 3 when $p_T^\gamma$ reaches 110 GeV$/c$.
%versus its transverse momentum $p_T^\gamma$ 
    Furthermore,  
    %it was shown 
     taking into account 
%    that by calculating this spectrum within perturbative QCD and using the    
     the CTEQ66c PDF,
     which includes the IC contribution obtained within the BHPS model \cite{Brodsky:1980pb,Brodsky:1981} 
     one can increase this ratio up to 1.5 
\cite{Stavreva:2010mw}. 
     For the $\gamma+b$-jets $p{\bar p}$-production no enhancement 
     in the $p_T^\gamma$-spectrum was observed 
     at the beginning of the experiment
\cite{D0:2009,Aaltonen:2009wc}.
     However in 2012 the D\O\ collaboration has confirmed observation of such an enhancement
\cite{Abazov:2012ea}.   

    This intriguing observation stimulates our interest 
    to look for a similar ``IC signal'' in $p p\rightarrow \gamma+c(b)+X$ 
    processes %with the ATLAS detector 
    at LHC energies.

\begin{figure}[h!]
\begin{center}
\epsfig{file=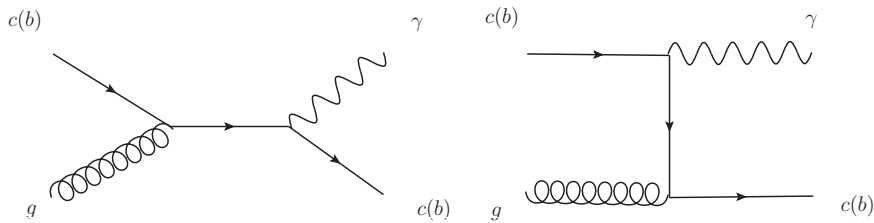,width=0.8\linewidth}
%\centerline{\includegraphics[width=0.50\textwidth]{gc_gammac.ps}}
 \caption{The Feynman diagrams for the hard process $c(b) g\rightarrow \gamma c(b)$, the one-quark exchange
  in the s-channel (left) and the same in the t-channel (right).
} 
\label{Fig_Fd_5}
\end{center}
\end{figure}

    The LO QCD Feynman diagrams for the process $c(b)+g\rightarrow\gamma+c(b)$  
    are presented in 
Fig.~\ref{Fig_Fd_5}.  

    These hard sub-processes give the main contribution to the reaction 
    $pp\rightarrow\gamma + c(b)$-jet$+X$.

    Within LO QCD, 
    in addition to the main subprocesses illustrated in 
Fig.~\ref{Fig_Fd_5}
    one considers the 
    subprocesses $gg\rightarrow c{\bar c}$, $q c\rightarrow q c$, $g c\rightarrow g c$ 
    accompanied by the bremstrallung $c({\bar c})\rightarrow c\gamma$,
    the contribution of which is sizable at low $p_T^\gamma$  and can be neglected at
    $p_T^\gamma>$ 60 GeV$/$c, according to \cite{Lipat-Zot:2012}. 
    The diagrams within the NLO QCD are more complicated than 
Fig.~\ref{Fig_Fd_5}. 

    Let us illustrate qualitatively the kinematical regions where 
    the IC component can contribute significantly 
    to the spectrum of prompt photons produced together with a $c$-jet in $pp$ collisions at the LHC. 
    For simplicity we consider only the contribution 
    to the reaction $pp\rightarrow\gamma+c(jet)+X$
    of the diagrams given in 
Fig.~\ref{Fig_Fd_5}. 
    According to (7) and (\ref{def:xijmin}),
    at certain values of the transverse momentum of the photon, $p_T^\gamma$, 
    and its pseudo-rapidity, $\eta_\gamma$, (or rapidity $y_\gamma$)
    the momentum fraction of $\gamma$ %((\ref{def:xFetapt})) 
    can be $x_{F \gamma}>0.1$, 
    therefore the fraction of the initial $c$-quark must %    (Fig.~\ref{Fig_Fd_5}) 
    also be above 0.1, where the IC contribution in the proton is enhanced 
(see Fig.~\ref{Fig_2IC}).    
    Therefore, one can expect some non-zero IC signal in the $p_T^\gamma$ 
    spectrum of the reaction $pp\rightarrow\gamma+c+X$ in this
    certain region of $p_T^\gamma$ and $y_\gamma$. 
    In principle, a similar qualitative IC effect can be visible in the production
    of $\gamma^*/Z^0$ decaying into dileptons accompanied by $c$-jets in $pp$ collisions.   

    Experimentally one can measure the prompt photons accompanied by the $c(b)$-jet corresponded
    to the hard subprocess  $c(b) g\rightarrow \gamma c(b)$ presented in Fig.~\ref{Fig_Fd_5}, when 
    $\gamma$ and $c(b)$-jet are emitted back to back. Therefore, it would be interesting 
    to look the contribution of this graph to the $p_T^\gamma$ spectrum compared to total QCD calculation 
    including the NLO corrections.    

\begin{figure}[h!]
\begin{center}
\epsfig{file=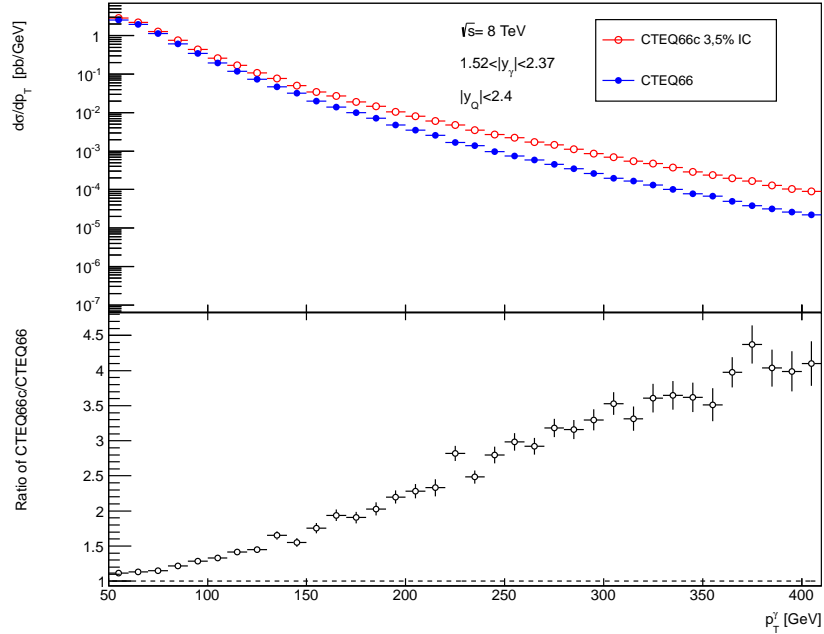,width=0.8\linewidth}
%\centerline{\includegraphics[width=0.50\textwidth]{FinGammaPt_Y1p52_Y2p37_Pt400.ps}}
 \caption{The distribution $d\sigma/dp_T^\gamma $  
of prompt photons produced in the reaction 
$pp\rightarrow\gamma c X$ over the transverse momentum $p_T^\gamma$
integrated over $dy$ in the interval 1.52$<\mid y_\gamma\mid$ 2.37, $<\mid y_c\mid<$ 2.4 at 
$\sqrt{s}=$ 8 TeV. The red open points correspond to the inclusion of
the IC contribution in PDF CTEQ66c with the IC probability of about 3.5\% 
\cite{Nadolsky:2008zw};
the blue solid points is our calculation using the CTEQ66 without the IC 
contribution in the proton. 
The calculation was done within PYTHIA8 using the LO QCD and including the ISR, FSR
and MPI.
}
\label{Fig_diffcrs_2.0}
\end{center}
\end{figure}

In Fig.~\ref{Fig_diffcrs_2.0} the distribution $d \sigma/dp_T^\gamma$ of prompt photons 
       produced in the reaction $pp\rightarrow\gamma+c+X$ at $\sqrt{s}=8$~TeV is presented 
       in the interval of the photon rapidity $1.52<\mid y_\gamma\mid<2.37$ and the $c$-jet rapidity
       $\mid y_c\mid< 2.4$.
       The calculation was carried out within PYTHIA8 \cite{PYTHIA8} 
%and only graphs in Fig.~\ref{Fig_Fd_5} 
      including only graphs in Fig.~\ref{Fig_Fd_5}
      and the radiation corrections for the initial (ISR) 
      and final (FSR) states along with the multi-parton 
      interactions (MPI) within PYTHIA8.

       The upper line in the top of 
Fig.~\ref{Fig_diffcrs_2.0} is calculated with the CTEQ66c PDF and includes IC,  
       while the lower line uses the CTEQ66 PDF where the charm PDF is radiatively generated only.
       The probability of the IC contribution used is about 3.5\%    
\cite{Nadolsky:2008zw}, this yields the highest sensitivity of the cross-section to the IC, however the 
       intrinsic charm 
       in the proton could also be about 1\% \cite{Pumplin:2007wg} and therefore the results  in this case 
       will yield a lesser difference when compared to the radiatively generated ones. 
       The ratio of the spectra with IC and without IC as a function of $p_T^\gamma$ is 
       presented in the bottom of Fig.~\ref{Fig_diffcrs_2.0}. 
   
       One can see from 
Fig.~\ref{Fig_diffcrs_2.0} that the inclusion of the IC contribution 
       increases the spectrum by a factor of 4-4.5 at $p_T^\gamma \simeq 400$ GeV$/c$, 
       however the cross section is too small here (about 1 fb). 
       At $p_T^\gamma\simeq$ 150-200 GeV$/c$ the cross section is about 8--30 fb
       if the IC is included and the IC signal reaches 250\% - 300\%. 
       It corresponds to 800--3000 events in the 5 GeV$/$c bin for the luminosity $L = 20$ fb$^{-1}$.

       Naturally the $p_T^\gamma$ distribution in 
Fig.~\ref{Fig_diffcrs_2.0} has the same form as the distribution over
       the transverse momentum of the $c$-quark,  $p_T^c$, 
       when only the hard subprocess $g+c\rightarrow\gamma+c$ in 
Fig.~\ref{Fig_Fd_5} is included.  

      According to this figure, the IC signal can be about 180\%-250\% 
      at $p_T^\gamma\simeq$ 150-200 GeV$/$c and the cross section
       is about 10--40 fb, which corresponds 
       to about 1000--4000 events in the 5 GeV$/$c bin at $L$=20 fb$^{-1}$.
       The NLO QCD calculations showed the similar results \cite{BDLSS:2013}.
\subsection{$W$-boson and $b$-jet production}

\begin{figure}[h!]
\begin{center}
\epsfig{file=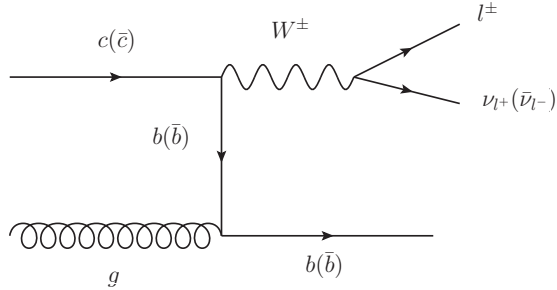,width=0.5\linewidth}
%\centerline{\includegraphics[width=0.350\textwidth]{cg_W_b.pdf}}
\caption{The Feynman diagrams for the hard process $c g\rightarrow W^\pm b$, the one-quark exchange
  in the t-channel.
}
\label{Fig_cg_Wb}
\end{center}
\end{figure}

\begin{figure}[h!]
\begin{center}
\epsfig{file=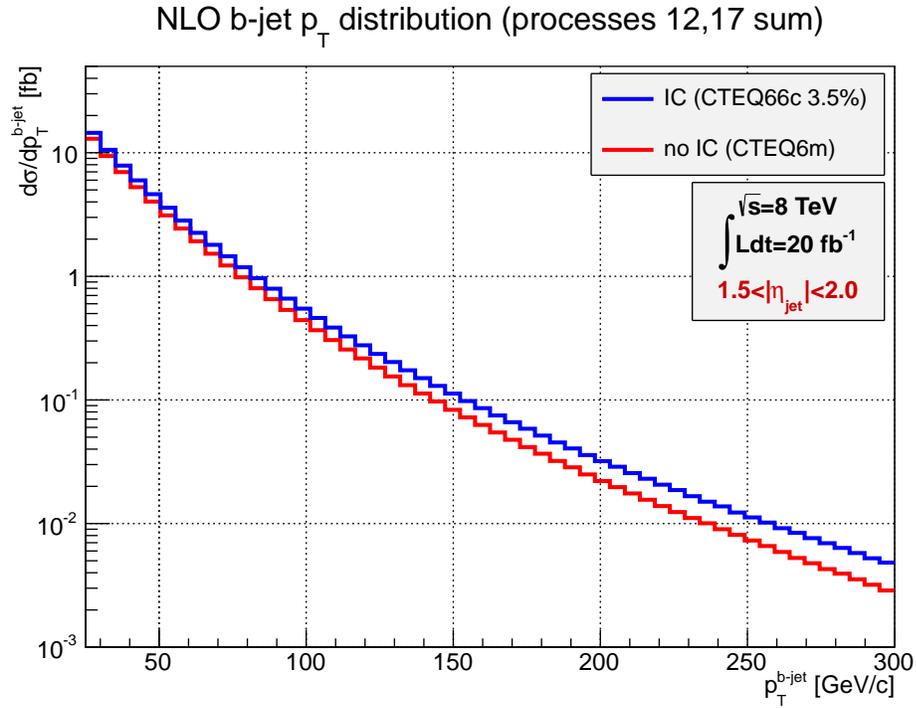,width=0.8\linewidth}
%\centerline{\includegraphics[width=0.40\textwidth]{b-jet-pt-15-eta-20.ps}}
 \caption{The transverse momentum spectrum of $W^\pm$-boson accompanied by the $b$-jet
produced in $pp$ collision at the LHC energy $\sqrt{s}=$ 8 TeV
and at the $b$-jet pseudo-rapidity 1.5$<\eta_b<$2. The down line corresponds to
the calculation without the IC, the upper curve is our result including the IC contribution 
in the PDF CTEQ6,6c with the probability about 3.5 $\%$. 
}
\label{Fig_IC_b}
\end{center}
\end{figure}

\begin{figure}[h!]
\begin{center}
\epsfig{file=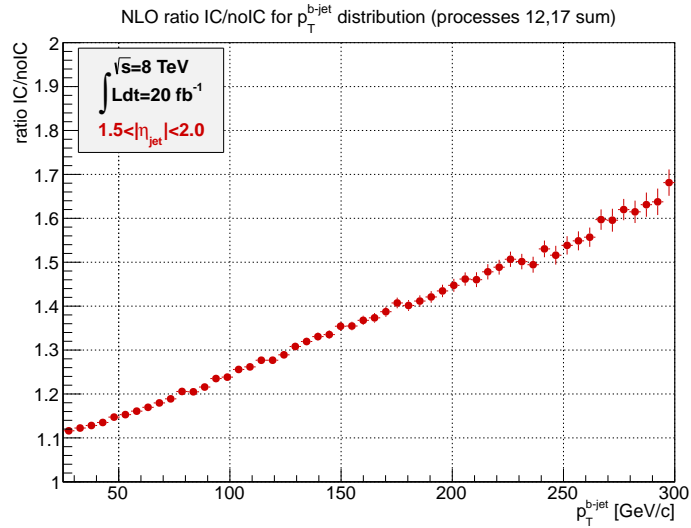,width=0.6\linewidth}
%\centerline{\includegraphics[width=0.40\textwidth]{b-jet-pt-15-eta-20-ratio.pdf}}
 \caption{The ratio of spectra presented in Fig.~\ref{Fig_IC_b} with  and without IC contribution
in the proton.  
} 
\label{Fig_IC_b_ratio}
\end{center}
\end{figure}

Let us analyze now another process, the production of vector boson accompanied by the $b$-jet 
in $pp$ collision. The LO QCD diagram for the process $c({\bar c})+g\rightarrow W^\pm +b({\bar b})$ 
is presented in Fig.~\ref{Fig_cg_Wb}. These hard subprocesses can give the main contribution to the 
reaction $pp\rightarrow W^\pm(\rightarrow l^+ + l^-)+b(\bar b)-jet +X$, which could give us also 
the information on the IC contribution in the proton.  
In Fig.~\ref{Fig_IC_b} the transverse momentum spectrum of $W^+$-boson accompanied by the $b$-jet
produced in $pp$ collision at the LHC energy $\sqrt{s}=$ 8 TeV is presented. 
The calculation was done within the MCFM generator \cite{MCFM:2013}.
The down line corresponds 
to the calculation without the IC, the upper curve is our result including the IC contribution 
in the PDF CTEQ6,6c with the probability about 3.5 $\%$ 
In Fig.~\ref{Fig_IC_b_ratio} the ratio of spectra presented in Fig.~\ref{Fig_IC_b} with  
and without IC contribution in the proton is presented.
In Figs.~(\ref{Fig_IC_b},\ref{Fig_IC_b_ratio}) the sum of the $p_T^{b-jet}$ spectra for the processes
$pp\rightarrow W^+(\rightarrow e^+ + \nu) + {\bar b} + X$ (process 12) and
$pp\rightarrow W^-(\rightarrow e^- + {\bar\nu}) + b + X$ (process 17) is presented. 
One can see from Figs.~(\ref{Fig_IC_b},\ref{Fig_IC_b_ratio}) that the inclusion of the IC contribution in the
PDF results in the increase of the transverse momentum spectrum of $W$-bosons by a factor of 2 at 
$p_T^{b-jet}~$ 250-300 GeV$/$c.  
%\begin{figure}[h!]
%\epsfig{file=cg_W_b.eps,width=0.8\linewidth}
%%\centerline{\includegraphics[width=0.350\textwidth]{cg_W_b.pdf}}
%\caption{The Feynman diagrams for the hard process $c g\rightarrow W^\pm b$, the one-quark exchange
%  in the t-channel (right)
%}.
%\label{Fig_cg_Wb_2.5}
%\end{figure}
\begin{figure}[h!]
\begin{center}
\epsfig{file=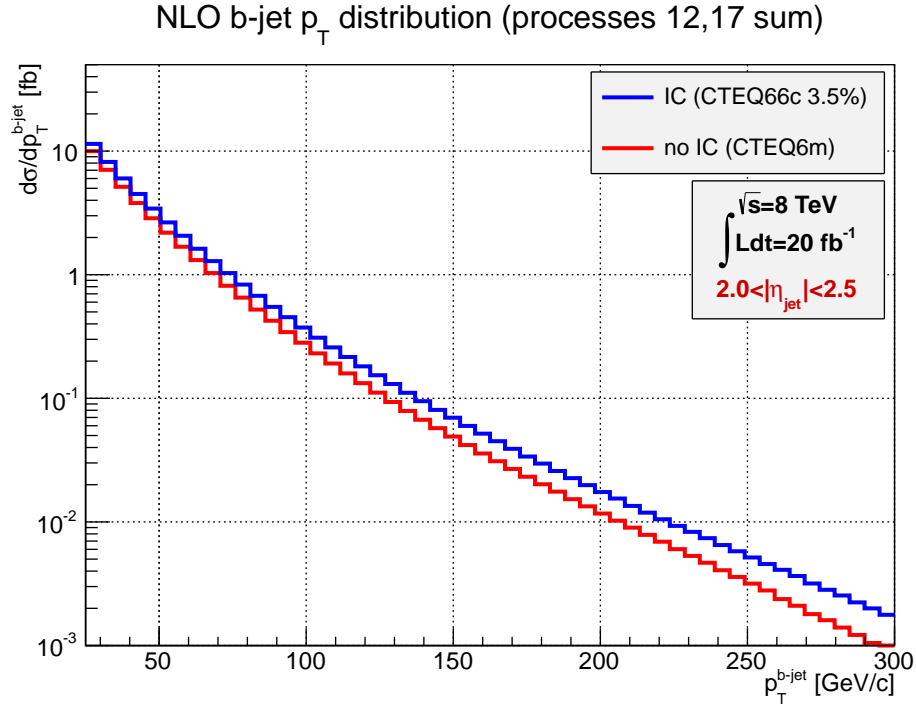,width=0.8\linewidth}
%\centerline{\includegraphics[width=0.40\textwidth]{b-jet-pt-15-eta-20.ps}}
 \caption{The transverse momentum spectrum of $W^\pm$-boson accompanied by the $b$-jet
produced in $pp$ collision at the LHC energy $\sqrt{s}=$ 8 TeV and 
at the $b$-jet pseudo-rapidity 2.0$<\eta_b<$2.5. The down line corresponds to
the calculation without the IC, the upper curve is our result including the IC contribution 
in the PDF CTEQ6,6c with the probability about 3.5 $\%$. 
} 
\label{Fig_IC_b_2.5}
\end{center}
\end{figure}
\begin{figure}[h!]
\begin{center}
\epsfig{file=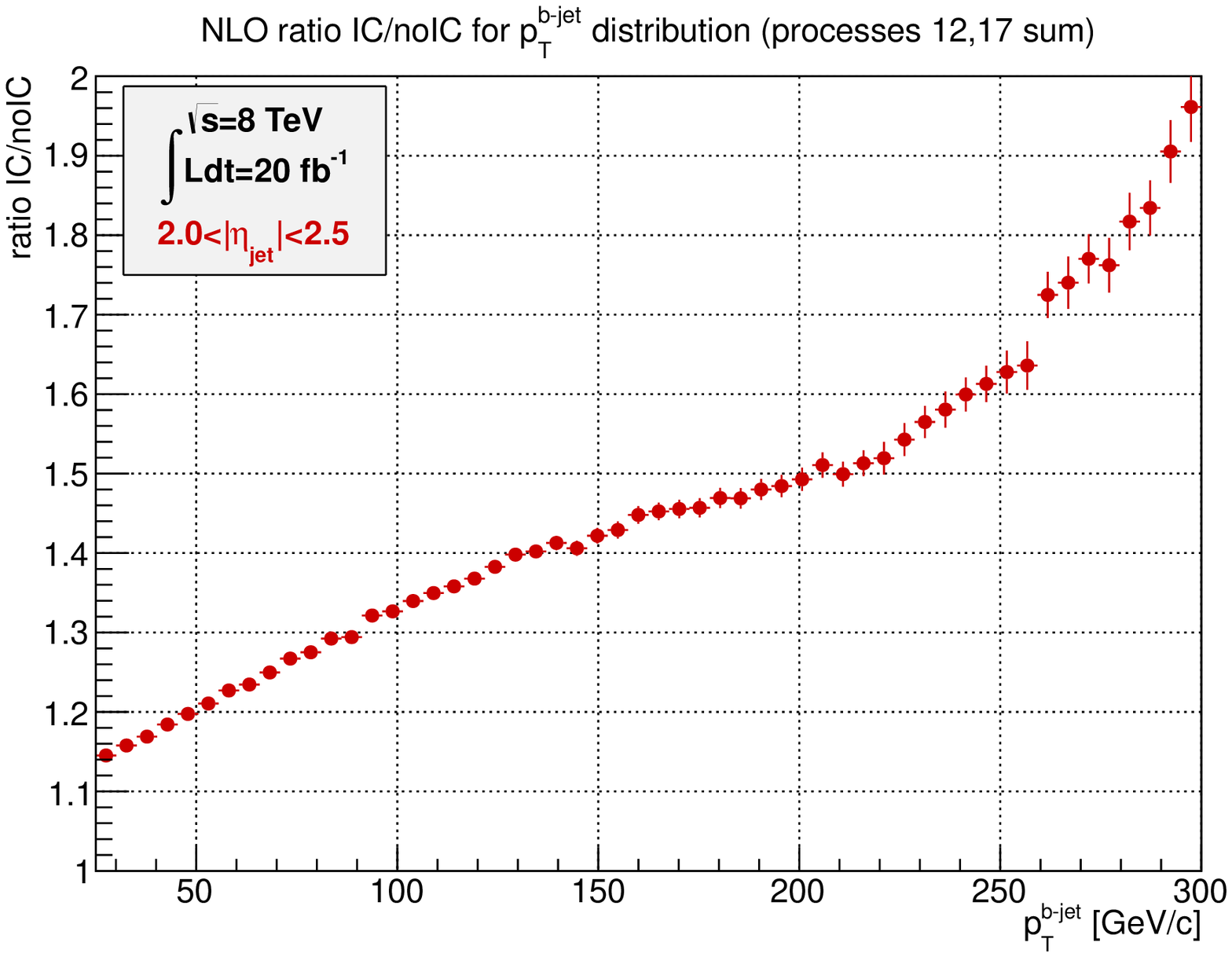,width=0.6\linewidth}
%\centerline{\includegraphics[width=0.40\textwidth]{b-jet-pt-15-eta-20-ratio.pdf}}
 \caption{The ratio of spectra presented in Fig.~\ref{Fig_IC_b_2.5} with  and without IC contribution
in the proton.  
}
\label{Fig_IC_b_ratio_2.5}
\end{center}
\end{figure}
%%%%%%%%%%%%%%%%%%%%%%%%%%%%%%%%%%%%%%%                      %%%%%%%%%%%%%%%%%%%%%%%%%%%%%%%%%%%%%%%%%%%%%%%%%%%%%
\section{Conclusion} 

%%%%%%%%%%%%%%%%%%%%%%%%%%%%%%%%%%%%%%%%%%%%%%%%%%%%%%%%%%%%%%%%%%%%%%%%

          We analyzed the inclusive $K^-$-meson production in $pp$ collision at the initial energy 
          $E_p=$158 GeV and gave some predictions for the NA61 experiment going on at CERN. We showed that
          in the inclusive spectrum of $K^-$-mesons as a function of $p_t$ 
          at some values of their rapidities
          the signal of the {\it intrinsic} strangeness can be visible and reach about 200$\%$ and more at large 
          momentum transfer we took.      
          The probability of the {\it intrinsic} strangeness to be about 2.5$\%$, as 
          was found from the best description of the HERA and HERMES data on the DIS, see \cite{IS:2012} and
          references therein.
          The similar predictions can be made for the open strangeness production at the energies of the 
          CBM (Darmstadt) and NICA (Dubna) experiments. The main goal of such predictions is to show that 
          at the certain kinematical region the contribution of the {\it intrinsic} strangeness in the proton
          can result in the enhancement in inclusive spectra by a factor of 2-3. 

       In this paper we also have shown 
      that the possible existence of the intrinsic heavy quark components in the proton
      can be seen not only in the forward open heavy flavor production in $pp$-collisions
      (as it was believed before) but it can be visible also 
      in the semi-inclusive $pp$-production of prompt photons and $c$-jets
      at rapidities 1.5$<\mid y_\gamma\mid<$ 2.4, $\mid y_c\mid<2.4$ 
      and large transverse momenta of photons and jets.

      In the inclusive photon spectrum measured together with a c-jet 
      a rather visible enhancement can appear due to the intrinsic charm  (IC) quark
      contribution. 
      In particular, it was shown that the IC contribution  
      can produce much more events (factor 2 or 3) at $p^\gamma_{T}>$ 150 GeV$/$c 
      and forward $y_\gamma$ 
      in comparison with the relevant 
      number expected in the absence of the IC. 
      Furthermore the same enhancement is also coherently expected in the  
      transverse momentum, $p_T^c$, distribution of the $c$-jet measured together with 
      the above-mentioned prompt photon in the $p p\rightarrow\gamma + c$-jet$+X$
      process.    
      
      As we have shown also, the possible signal of {\it intrinsic} charm could be observed in the 
      production of vector mesons $W^\pm$ accompanied by the $b$-jet. Our prediction on the IC contribution
      in the process $pp\rightarrow W^\pm + b-jet + X$ is similar to to the one for the reaction 
      $pp\rightarrow \gamma + c-jet + X$         
      
\section{Acknowledgements}
%We are very grateful to A.F Pikelner 
%for his help with the MC calculations.
We thank S.J. Brodsky,  M. Gazdzicki, A. Glazov, S.M. Pulawski and  A. Rustamov
for extremely helpful discussions and recommendations for the 
predictions on the search for the possible intrinsic heavy flavour components in
$pp$ collisions at high energies. G.L. thanks the coauthors V.A. Bednyakov, T. Stavreva 
and M. Stockton of our common paper \cite{BDLSS:2013} for very fruitful collaboration.  
We are also grateful to S.P. Baranov, H. Beauchemin, A.M. Cooper-Sarkar, H. Jung, B. Kniehl, 
B.Z. Kopeliovich, A. Likhoded, A.V. Lipatov, E. Meoni, M. Poghosyan, A. Romsoyan, 
V. Radescu, P. Spradlin, D. Stump, V.V. Uzhinsky and N.P. Zotov 
for very helpful discussions. 
This work was supported in part by the Russian Foundation for Basic Research, 
grant No: 11-02-01538-a and No: 13-02001060.

%To acknowledge funding bodies etc., a special section may be placed
%before the bibliography: \verb?\section*{Acknowledgements}?.
% ****************************************************************************
% BIBLIOGRAPHY AREA
% ****************************************************************************
% please do not change the following line
%\begin{footnotesize}
%\bibliographystyle{eplbib} 
%\bibliography{ic}
%\end{footnotesize}
\begin{footnotesize}
%\begin{thebibliography}{00}

\end{footnotesize}
%%%%%%%%%%%%%%%%%%%%%%%%%%%%%%%%%%%%%%%%   %%%%%%%%%%%%%%%%%%%%%%%%%%%%%%%%%%%%%%%%%%%%%%
%%%%%%%%%%%%%%%%%%%%%%%%%%%%%%%%%%%%%%%%   %%%%%%%%%%%%%%%%%%%%%%%%%%%%%%%%%%%%%%%%%%%%%%

%% The Appendices part is started with the command \appendix;
%% appendix sections are then done as normal sections
%% \appendix

\end{document}